\documentclass{Interspeech2024}
\usepackage{amsmath,graphicx,bm,amsfonts,mathtools}
\usepackage{color}
\usepackage{multirow, tabu}
\usepackage{caption, hyperref, enumitem, setspace}
\usepackage{lipsum}

\interspeechcameraready

\title{Multimodal Representation Loss Between Timed Text and Audio\\for Regularized Speech Separation}

\name[affiliation={1}]{Tsun-An}{Hsieh}
\name[affiliation={2}]{Heeyoul}{Choi}
\name[affiliation={1}]{Minje}{Kim}

\address{
  $^1$Department of Computer Science, University of Illinois at Urbana-Champaign, IL, USA\\
  $^2$School of Computer Science and Electrical Engineering, Handong Global University, Korea
}
\email{tsunanh2@illinois.edu, hchoi@handong.edu, minje@illinois.edu}


\begin{document}

\maketitle

\begin{abstract}
    Recent studies highlight the potential of textual modalities in conditioning the speech separation model's inference process. However, regularization-based methods remain underexplored despite their advantages of not requiring auxiliary text data during the test time. To address this gap, we introduce a timed text-based regularization (TTR) method that uses language model-derived semantics to improve speech separation models. Our approach involves two steps. We begin with two pretrained audio and language models, WavLM and BERT, respectively. Then, a Transformer-based audio summarizer is learned to align the audio and word embeddings and to minimize their gap. The summarizer Transformer, incorporated as a regularizer, promotes the separated sources' alignment with the semantics from the timed text. Experimental results show that the proposed TTR method consistently improves the various objective metrics of the separation results over the unregularized baselines.
\end{abstract}

\section{Introduction}
\label{sec:intro}

Recent studies have shown significant progress in deep learning-based audio source separation \cite{ChenZ2017deep,ZeghidourN2020wavesplit, HersheyJ2016icassp, StollerD2018waveunet, DefossezA2019mss, TzinisE2020sudormrf}, among which end-to-end approaches are popular approaches. For example, Conv-TasNet \cite{LuoY2019conv-tasnet} established the encoder-separator-decoder structure, which removed traditional time-frequency feature extraction, such as magnitude spectrogram or mel-frequency cepstral coefficients. Dual-Path RNN \cite{YiL2020dualpathRNN} followed to capture both the temporal and spatial dependencies through modeling across both directions. In addition, introducing the Transformer \cite{VaswaniA2017transformer} architectures to source separation also advanced the performance due to their self-attention mechanism, such as Dual-Path Transformer \cite{Chen2020dual} and SepFormer \cite{Subakan2021attention, SubakanC2023sepformer}. 

% Innovative loss functions have also contributed to the advancements. Permutation invariant training (PIT) \cite{YuD2017pit}, for example, successfully resolved the inherent permutation issue when the model computes the source-specific reconstruction loss. Mixture invariant training \cite{WisdomS2020mixit} relaxed the separation problem by re-defining the input as a mixture of mixtures. 

Another category of studies focuses on leveraging cross modality clues, e.g. visual and textual queries, as auxiliary information that conditions the separation system. Hence, this type of systems is suitable for extracting out a source of interest defined by the cue, which is a task often called \textit{target source extraction} (TSE). Query-based approaches have been widely investigated in the fields of singing voice separation \cite{schulze2021phoneme}, speech separation \cite{schulze2020joint}, and sound separation \cite{kilgour2022text,liu2022separate,Dong2022CLIPSepLT,tzinis2022heterogeneous}. Among these works, the text modality has been one of the primary ways to convey information about the target source. For example, in \cite{kilgour2022text}, textual description or sample audio from the same speaker is used to designate the target speaker; LASS-Net \cite{liu2022separate} conditions the hidden vectors in the separation network with a Transformer-based query network to extract textually described sounds; CLIPSep \cite{Dong2022CLIPSepLT} used contrastive language-image-audio pretraining to learn a joint embedding for trimodal representation and used it for TSE; in \cite{tzinis2022heterogeneous}, a heterogeneous TSE task is defined to condition the separation network using various concepts such as gender, language, and loudness.

Although these successful TSE systems use auxiliary information for effective conditioning, using other modalities to \textit{regularize} the separation task during training has not been studied in depth. The Voice ID loss \cite{Shon2019VoiceID_loss} is an example, where the speech enhancement model is regularized to reduce both the typical reconstruction and the speaker verification loss. The prototypical speaker-interference (PSI) loss \cite{Mun2022interspeech} also uses a speaker-level representation loss to regularize a TSE system.
While these regularization methods are effective, they are limited to the audio modality, leaving room for investigating other modalities, such as text. For example,
using an ASR loss jointly with the signal reconstruction loss is another promising multimodal approach while they are more focused on the ASR performance than reconstruction and do not consider inter-word relationships as in our method \cite{von2020end,Shi2022joint,yuma2022snri}. 
\begin{figure}[t!]
  \centering
  \centerline{\includegraphics[width=\linewidth]{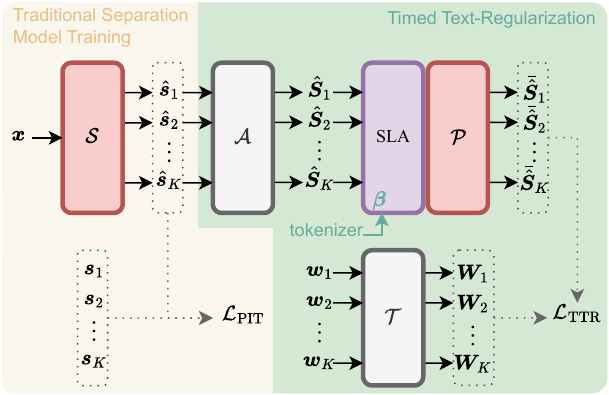}}
  \caption{The proposed TTR-SS training pipeline. The yellow and green areas represent the traditional training of a speech separation model and the proposed TTR, respectively. The test-time inference only uses the yellow area.}
  \label{fig:joint}
  % \vspace{-1pt}
\end{figure}

% The simplest idea to achieve this goal is to simply match the acoustic features with the corresponding word embeddings. However, this is challenging becuase direct comparison between feature vectors in different domains is infeasible in practice since the information entailed in acoustic features could largely differ from that in word embeddings, and training a complicated large mapping function to transform the acoustic features to their target word embeddings is also unrealistic. 
In this work, our main contribution is the multimodal representation loss defined between timed text and speech audio to regularize speech separation models. Compared to the other TSE methods, where the separation model is indirectly conditioned by auxiliary text \cite{kilgour2022text, liu2022separate} or an unaligned script \cite{schulze2021phoneme}, our text modality is high-quality and strongly associated with the source audio. In particular, we use source-specific scripts of all the clean utterances for training, assuming their word-level synchronization with the audio frames. This assumption is rather strong to condition the model during testing as in the TSE methods because acquiring a time-aligned transcript of the test-time sources is extremely difficult. On the contrary, the proposed TTR method provides consistent performance improvement at no additional cost, i.e., without asking the user for a time-aligned transcript of the target source.  Furthermore, since the proposed method is to regularize the training objectives, it does not add any computational cost to the test-time inference. 

Figure \ref{fig:joint} illustrates the proposed regularization method for speech separation. First, the two pretrained audio and text encoders, $\mathcal{A}$ and $\mathcal{T}$, convert $K$ separated speech sources $\hat{\bm s}_{k\in\{1,\ldots,K\}}$ and their corresponding timed text $\bm w_k$ into frame-level audio embeddings $\hat{\bm S}_k$ and subword-level text embeddings $\bm W_k$, respectively. Second, the subword-level alignment (SLA) module associates consecutive audio embeddings with a subword embedding. Third, the summarizer Transformer $\mathcal{P}$ follows to convert each subword-specific audio embeddings into a summary vector, representing the subword in the audio modality. Finally, our TTR loss computes the similarity between the subword-level aggregation of the audio embeddings $\bar{\hat{\bm S}}$ and the corresponding subword embedding from the text encoder.

% We pretrain the $\mathcal{P}$ network by minimizing the difference of the pairs of time-aligned embeddings. This way, the second-stage Transformer encoder learns to match the global relationship between the audio embeddings and the text embeddings. More details are provided in Sec. \ref{sec:format}.

% The major challenges in this study are: 
% \begin{itemize}
%     \item Audio embeddings represent phonetic information, while word embeddings entail semantics.
%     \item Given the forced word-level alignment, audio embeddings still require further process to be aligned with sub-words which commonly used by the tokenizers of language models.
% \end{itemize}\par
% To address the first issue, we introduce a self-similarity matching mechanism that forces the estimated sources to have similar intra-sequence dynamics as the paired word embeddings to escape from direct audio-to-word embedding mapping. For the second problem, a two-staged network is designed to align the audio embeddings with word embeddings. The first stage sub-network summarizes audio embeddings corresponding to a sub-word into a single feature vector. Next, the second stage sub-network learns to transform the summarized audio embeddings to have the similar internal dynamics of the corresponding word embeddings according to the proposed self-similarity matching score. 
% 

% 

% 
The pretrained summarizer, WavLM, and BERT, are frozen and combined as a regularization network, which provides an audio-text matching score to finetune the speech separation network. We call the finetuned separation network {\it Timed Text-Regularized Speech Separation} (TTR-SS).
Experimental results demonstrate that the proposed TTR-SS improves the performance of two baseline separation systems on two and three-speaker speech separation tasks with additive noise, perhaps due to the sentence-level semantics introduced to the loss function. Moreover, we note that TTR enhances the more complex SepFormer to a greater extent than it does the simpler Conv-TasNet, indicating that the TTR loss introduces more information for the network to learn from, requiring a larger model capacity. 

\section{Timed Text-Regularized Source Separation}
\label{sec:format}

% Let's use this formula:
% \begin{align}
%     \bm{X}\leftarrow \mathcal{T}(\bm{x})\\ %BERT, x is the raw word sequence
%     \bm{S}\leftarrow \mathcal{A}(\bm{s})\\ %wave2vec, s is the raw audio signal
%     \bm{Y}\leftarrow \mathcal{P}(\bm{S}) %projection
% \end{align}

\subsection{Problem Definition and the Loss Function}
Given a time-domain mixture signal $\bm{x}$ that consists of $K$ speech sources and a non-speech source $\bm{n}$, i.e., $\bm{x}=\sum_{k=1}^K \bm{s}_k + \bm{n}$, a speech separation model $\mathcal{S}$ is expected to estimate the sources back, $\{\hat{\bm{s}}_1, \hat{\bm{s}}_2, \dots, \hat{\bm{s}}_K\} \leftarrow \mathcal{S}(\bm{x})$, with a potential permutation of the source order. To compare each estimate to the best-matching target source, during training, the permutation invariant training (PIT) scheme \cite{YuD2017pit} is commonly used. In particular, 
% that consists of a permutation module and a similarity score function $g$, such as SI-SDR or scale-invariant source-to-noise ratio (SI-SNR) \cite{LuoY2018tasnet}. 
given a set of source estimates and a reconstruction loss function, e.g., negative SI-SDR, PIT searches for the best permutation that minimizes the total loss out of $K!$ potential permutations, whose $l$-th permutation is defined by $\bm{c}^{(l)}=\{i_1^{(l)},\ldots,i_K^{(l)}\}$:
\begin{equation}\label{eq:pit}
\mathcal{L}_\text{PIT}\coloneqq\min_{\bm{c}^{(l)},~~l\in\{1,\ldots, K!\}}\sum_{k=1}^K -\text{SI-SDR}\left(\bm{s}_k||\hat{\bm{s}}_{i_k^{(l)}}\right).
\end{equation}
\begin{figure}[t!]
  \centering
  \centerline{\includegraphics[width=\linewidth]{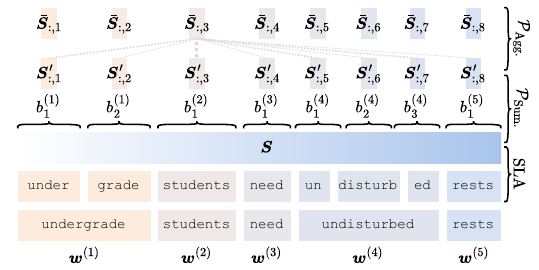}}
  \caption{Data flow of the SLA and summerizer Transformer. SLA divides audio embeddings into segments using subword boundaries. The $\mathcal{P}_{\rm Sum.}$ summarizes each segment into a single vector. $\mathcal{P}_{\rm Agg.}$ aggregates the summarized vectors to mimic the internal dependency of subword embeddings.}
  \label{fig:sst}
  % \vspace{-1pt}
\end{figure}

% $\in \mathbb{R}^{K\times K}$ that maximizes $g$, hence $\mathcal{L}_{\rm PIT} = -g(\bm{M}^*[\{\hat{\bm{s}}_1; \hat{\bm{s}}_2; \dots; \hat{\bm{s}}_K], [\{\bm{s}_1; \bm{s}_2; \dots; \bm{s}_K])$. \par
% \begin{align}
%     \bm{M} = \arg\max_{\bm{M}} g(\bm{M}\hat{\bm{S}}, \bm{S})
% \end{align}
\subsection{Subword-Level Alignment}
\label{ssec:ttr_ssm}
Timed text-regularization (TTR) takes advantage of the sentence-level semantics to guide the training of the separation model. To achieve this, we use a pretrained audio encoder and a language model (LM) to extract audio and word embeddings and then compare them to compute a regularization loss. We denote the extraction processes for text and audio by $\bm{W} = \mathcal{T}(\bm{w})$ and $\bm{S} = \mathcal{A}(\bm{s})$, respectively. Their input $\bm{w}$ and $\bm{s}$ are a sequence of subwords and a waveform, while $\bm{W}\in\mathbb{R}^{D_{\bm{W}}\times M}$ and $\bm{S}\in\mathbb{R}^{D_{\bm{S}}\times T}$ represent the word and audio embeddings that are $D_{\bm{W}}$ and $D_{\bm{S}}$-dimensional vectors, respectively. Note that $M$ is the number of subwords, which is usually much smaller than audio frames $N$.

We assume that reliable word boundaries of a ground-truth source utterance $\bm{s}$ are available, while their subword boundaries are not. Since a typical $\mathcal{T}$ model, e.g., BERT \cite{DevlinJ2019bert}, uses tokenized subwords as its atomic input elements, the resulting embedding vectors $\bm{W}$ are at the subword level, which the audio embeddings $\bm{S}$ should be aligned to. As a remedy, we propose the \textit{subword-level alignment} (SLA) algorithm (Figure \ref{fig:joint}), which infers the subword boundaries from the known word lengths. First, we assume the $l$-th word consists of $m_l$ subwords, i.e., $\bm{w}^{(l)}\!=\![w^{(l)}_1, \ldots, w^{(l)}_{m_l}]$. Hence, the entire subword sequence $\bm{w}\!=\![\bm{w}^{(1)}, \ldots, \bm{w}^{(L)}]=[w^{(1)}_1, \ldots, w^{(1)}_{m_1}, w^{(2)}_1, \ldots, w^{(2)}_{m_2}, \ldots w^{(L)}_1, \ldots, w^{(L)}_{m_L}]$, where $L$ denotes the total number of words in $\bm{w}$.
Therefore, $M\!=\!\sum_{l=1}^L m_l$.
Then, for the known word lengths $\bm{b}\!=\![b^{(1)}, \ldots, b^{(L)}]$, we simply postulate that each word length can be evenly divided, i.e., $b^{(l)}\!=\!b^{(l)}_1\!+\!\cdots\!+\!b^{(l)}_{m_l}$ and $b^{(l)}_m\!=\!b^{(l)}/m_l, ~\forall m$. The subdivision can redefine $\bm{b}\!=\![b^{(1)}_1, \ldots, b^{(1)}_{m_1}, b^{(2)}_1, \ldots, b^{(2)}_{m_2}, \ldots, b^{(L)}_1, \ldots, b^{(L)}_{m_L}]$ as the list of subword lengths. In addition, we also define the subword boundaries $\bm{\beta}\!=\![\beta_0,\beta_1,\ldots,\beta_{M-1}, \beta_M]$, where $\beta_m\!=\!\sum_{m'=1}^{m} \bm{b}_{m'}$, i.e., the sum of the first $m$ subword lengths, while $\beta_0=0$.
With this information, we can group the audio embeddings into subword-specific sub-sequences. For example, the $t$-th audio embedding $\bm{S}_{:,t}$ belongs to the $m$-th subword if $\beta_{m-1}\!<\! t/R \!<\! \beta_{m} $, where $R$ denotes the frame rate of the audio embedding, i.e., the number of embeddings per second.

% Next, as audio and word embeddings are unaligned, word-level alignment, which records the start and end time stamps of a word, is involved to line them up. 
% Based on the number of subwords in a word, where the subwords are derived from the subword tokenizer, which is the pre-processing step before the LM, the audio embeddings are split equally. Now, each subword corresponds to a specific audio embedding sequence, and we term this subword-audio embedding pair {\it subword-level alignment} (SLA). Details of subword-level alignment can be seen in Algorithm \ref{alg:swa}.
Specifically, BERT and WavLM \cite{chen2022wavlm} are adopted in this work for $\mathcal{T}$ and $\mathcal{A}$ for subword and audio embeddings extraction. The word-level boundaries are computed by Montreal Forced Aligner \cite{mcauliffe17_interspeech}, which are further divided into the subword level using the abovementioned SLA algorithm.
% 
% \begin{algorithm}[tb!]
%   \caption{Subword-Level Alignment}
%   \label{alg:swa}
% \begin{algorithmic}
%   \STATE {\bfseries 1.} \textbf{Inputs}: subword tokenizer $\mathcal{F}_{sub}$; word lengths $\bm{l}=[l_1, l_2, \dots, l_N]$; audio embedding $\bm{S}$; word sequence $\bm{w} = [w_1, w_2, \dots, w_N]$\\
%   \STATE {\bfseries 2.} \textbf{Aligning Subwords and Audio Embeddings}: \\
%   \FOR{$i$ from $1$ to $N$}
%     \STATE $\bm{tok}_i = [tok_{i1}, tok_{i2}, \dots, tok_{iM}] \gets \mathcal{F}_{sub}(\bm{w}_i)$ \# $M$ subwords in $i^{th}$ word\\
%     $\bm{l}'=[~]$
%     \FOR{$m$ from 1 to $M$}
%       % \IF{$l_i$ is divisible by $M$}
%       \STATE \# Equally split the word length in to $M$ segments.\\
%       \STATE Append $\frac{l_i}{M}$ to $\bm{l}'$ \\
%       % \ELSIF{$m=M$}
%       %   \STATE Append $l_i \% M$ to $\bm{l}'$ \\
%       % \ENDIF
%     \ENDFOR
%     \STATE $l_i \gets \bm{l}'$
%   \ENDFOR
%   \STATE Unfold $\bm{l}$ to obtain flattened sequence of subword lengths.\\
%   $\bm{S}' = [~]$, $st=0$
%   \\
%   \# Align audio embeddings with word embeddings using subword lengths.\\
%   \FOR{$l_i$ in $\bm{l}$}
%     \STATE Append $\bm{S}[st:st+l_i]$ to $\bm{S}'$\\
%     \STATE $s \gets s+l_i$
%   \ENDFOR
%   \RETURN $\bm{S}'$
% \end{algorithmic}
% \end{algorithm}
% % 
\subsection{Summarizer Transformer}
\label{ssec:second_stage}
A successful SLA results in multiple audio embeddings associated with a subword embedding due to the audio modality's higher frame rate. Hence a summarizing mechanism is required, for which we propose the \textit{summarizer Transformer} function $\mathcal{P}$. This Transformer consists of two parts: the subword-level summarizer $\mathcal{P}_{\rm Sum.}$ and the sentence-level aggregator $\mathcal{P}_{\rm Agg.}$ as shown in Figure \ref{fig:sst}. The subword summarizer, a two-layer Transformer encoder, summarizes each aligned audio embedding sequence within the subword boundary into a feature vector with the same dimensionality with a word embedding, i.e., $\bm{S}_{:,m}'\leftarrow\mathcal{P}_{\rm Sum.}(\bm{S}_{:,\beta_{m-1}\leq t<\beta_{m}})$. By repeating the process for all the subword-specific subsequences of $\bm{S}$, we get $\bm{S}'\in\mathbb{R}^{D_{\bm{W}}\times M}$. The sentence-level aggregator, which has an identical architecture to the summarizer, follows to transform the sequence of subword-level summarized audio embeddings into a final version, i.e., $\bar{\bm{S}}\leftarrow\mathcal{P}_{\rm Agg.}(\bm{S}')$, without changing the dimension and length, resulting in the final audio representation $\bar{\bm{S}}\in\mathbb{R}^{D_{\bm{W}}\times M}$. Note that the sentence-level aggregation is over the $M$ embeddings, ensuring that the final representation $\bar{\bm{S}}$ encodes long-term context across the entire sentence, which is otherwise missing during the subword-level summarizing process that operates within the subword boundaries.

Since the $\mathcal{P}$ function projects the audio embeddings $\bm{S}$ to the $D_{\bm{W}}$-dimensional space, where the audio embeddings are learned to be comparable to word embeddings.
% In addition to that, our more structured objective is to use semantic information from the language model, by capturing the internal dependencies among the subwords within a sentence. To this end, we propose to use the self-similarity matching (SSM) loss $\mathcal{L}_\text{SSM}$ to match internal dynamics of sequences from different modalities.
% Specifically, the SSM loss is defined by the mean squared error (MSE) of the self-similarity matrices of audio and word embeddings $C^\text{audio}$ and $C^\text{text}$:
% \begin{equation}
%     \mathcal{L}_\text{SSM} \coloneqq \text{MSE}(C^\text{audio}, C^\text{text}),
% \end{equation}
% where each entry in $C^{\text{audio}}_{i,j}$ and $C^{\text{text}}_{i,j}$ represents the cosine similarity of $\bar{\bar{\bm{S}}}_{:,i}$ and $\bar{\bar{\bm{S}}}_{:,j}$ and $\bm{W}_{:,i}$ and $\bm{W}_{:,j}$, respectively. 
The $\mathcal{P}$ function learned to minimize the following timed text-regularization (TTR) loss function:
\begin{align}
\label{eq:ttr}
    \mathcal{L}_\text{TTR}(\bar{\bm{S}}, \bm{W})
    \coloneqq \frac{1}{M}\sum_{m=1}^M
    \left(
    1 - \frac{\bar{\bm{S}}_{:,m}\cdot\bm{W}_{:,m}}{\Vert\bar{\bm{S}}_{:,m}\Vert \Vert\bm{W}_{:,m}\Vert}
    \right),
\end{align}
i.e., the mean of the cosine distance between each pair of aligned subword-level embeddings from both modalities. Note that we use clean speech sources to pretrain $\mathcal{P}$, which is then frozen during the training of the separation model $\mathcal{S}$.
\subsection{Timed Text-Regularized Source Separation}
\label{ssec:finetune}
% After the ST module is pretrained, we run it to compute the audio embeddings' subword-level summary and then compare it with the sentence semantics acquired by $\mathcal{T}$. The difference between its pretraining and finetuning for regularization is that, for the latter, ST takes source estimates $\hat{\bm{s}}$ as input rather than the clean utterances. Hence, the regularization loss generated by imperfect separation will be backpropagated to the separator module $\mathcal{S}$ to improve its separation performance. 

Figure \ref{fig:joint} shows the finetuning pipeline for TTR-SS. The yellow area shows the traditional data flow of a speech separation model $\mathcal{S}$, which we pretrain using an ordinary speech separation pipeline: it takes a mixture $\bm{x}$ as input and predicts their constituent speech sources in the optimal order with the help from the PIT loss as shown in eq. \eqref{eq:pit}. The timed-text regularizer (TTR) is also pre-trained as described in Sec. \ref{ssec:ttr_ssm} and \ref{ssec:second_stage}, using clean speech utterances and their corresponding timed texts. 

The finetuning step further updates the separation module $\mathcal{S}$, while the summarizer Transformer, BERT, and WavLM modules are kept frozen. The difference is that, for finetuning, $\mathcal{P}$ takes the audio embeddings extracted from the source estimates as input rather than the clean utterances, i.e., $\bar{\hat{\bm{S}}}_k\leftarrow\mathcal{P}(\mathrm{SLA}(\mathcal{A}(\hat{\bm{s}}_k))$. 
Finally, we jointly minimize the PIT and TTR losses that are defined as the following total loss function:
\begin{align}
    \mathcal{L}_\text{total}\coloneqq
    \sum_{k=1}^K 
    \mathcal{L}_\text{PIT}(\hat{\bm{s}}_k, \bm{s}_k) + \lambda\mathcal{L}_\text{TTR}(\bar{\hat{\bm{S}}}_k, \bm{W}_k),
\end{align}
with a blending weight $\lambda$ chosen from $\{0.1, 0.5, 1.0\}$.

\section{Experimental Setup}
\subsection{Dataset Description \& Evaluation Metrics}
We use the LibriMix dataset \cite{CosentinoJ2020librimix} to train and validate our proposed method. 
% is designed to be used for research on speech separation. It is intended to be a generalizable dataset that can be used to train models that are capable of separating speech signals in a wide range of real-world scenarios. In LibriMix dataset, only the \textit{train-clean-100}, \textit{train-clean-360}, \textit{dev-clean}, and \textit{test-clean} subsets of LibriSpeech \cite{PanayotovV2015Librispeech} are used, representing around 470 hours of speech from 1,252 speakers with a 60K vocabulary. 
We use four subsets, Libri2Mix-Clean, Libri3Mix-Clean, Libri2Mix-Noisy, and Libri3Mix-Noisy, that consist of clean two- and three-speaker mixtures or their noisy versions. The speech sources are derived from LibriSpeech \cite{PanayotovV2015Librispeech}, while the noisy mixtures use ambient noises from the WHAM! dataset \cite{WichernG2019wham}. LibriMix follows the same structure as WHAM! and has two training sets, one validation set, and one test set. 
% The speed of the training noise samples were perturbed with factors of 0.8 and 1.2 to cover the \textit{train-360} subset. 
In this work, we use the \textit{train-360}, \textit{dev}, and \textit{test} as the training, validation, and test sets, respectively, with an 8KHz sample rate. In addition, we test the systems on both the clean and noisy mixtures.
% \subsection{Evaluation Metrics}
Evaluation metrics used are the traditional BSS\_Eval toolbox's decomposition of source-to-distortion ratio (SDR) into source-to-interference ratio (SIR) \cite{VincentE2006ieeeaslp} as well as the scale-invariant SDR (SI-SDR) metric \cite{LeRouxJL2018sisdr} and the short-time objective intelligibility (STOI) score \cite{TaalC2010icassp}. 
% \begin{align}
%    \hat{s} = s_{\rm target} + e_{\rm interf} + e_{\rm noise} + e_{\rm artif}
% \end{align}
% where $s_{\rm target}$ is the clean, non-distorted source; $e_{\rm interf}$, $e_{\rm noise}$, and $e_{\rm artif}$ refer to the error for interference, noise, and artifacts, respectively. The SAR, SIR, SI-SDR scores are formulated by:
% \begin{align}
%     {\rm SAR} &\coloneqq 10\log_{10}\left(\frac{\Vert s_{\rm target} + e_{\rm interf} + e_{\rm noise}\Vert^2}{\Vert e_{\rm artif}^2\Vert}\right)\\
%     {\rm SIR} &\coloneqq 10\log_{10}\left(\frac{\Vert s_{\rm target} + e_{\rm interf} + e_{\rm noise}\Vert^2}{\Vert e_{\rm artif}^2\Vert}\right)\\
%     {\rm SAR} &\coloneqq 10\log_{10}\left(\frac{\Vert s_{\rm target} + e_{\rm interf} + e_{\rm noise}\Vert^2}{\Vert e_{\rm artif}^2\Vert}\right)\\
% \end{align}
\subsection{Model Architecture and Training Setup}
\noindent{\bf Baseline Model and Joint Finetuning~} We adopt Conv-TasNet \cite{LuoY2019conv-tasnet} and SepFormer \cite{Subakan2021attention, SubakanC2023sepformer} as our baseline models. Conv-TasNet consists of a 1-D convolutional encoder, a separator, and a decoder. The convolution encoder first encodes the raw waveforms into a 2-D feature map. The separator leverages the temporal convolution to estimate feature masks that separate the encoded 2-D feature map. Finally, the separated feature maps are transformed back to waveform predictions. Specifically, the 1-D convolutional encoder contains 24 convolutional blocks, where each block has 512 channels with a kernel size of 16, and each kernel strides by 8.
With a similar encoder-separator-decoder structure, SepFormer mainly relies on the Transformer-based dual-path processing blocks as the separator. SepFormer repeats the separator twice, which contains eight layers of Transformers for inter- and intra-paths, and each Transformer layer has eight attention heads.
The baseline model pretraining and TTR-SS finetuning share the same optimization configuration except that the loss functions are $\mathcal{L}_\text{PIT}$ and $\mathcal{L}_\text{total}$, respectively. We use the Adam optimizer \cite{KingmaD2015adam} with a learning rate of $10^{-3}$ and $1.5\times 10^{-4}$ for Conv-TasNet and SepFormer, respectively. A learning rate scheduler halves the current learning rate if the validation loss is not reduced for five epochs. The training and finetuning are early-stopped if no improvement is seen after 30 epochs.

\noindent{\bf WavLM and BERT} Two pre-trained models, WavLM and BERT, extract audio and word embeddings that contain phonetic and semantic information, respectively. Both use a Transformer encoder-based architecture, and we used their publicly available pretrained versions. To minimize computational overhead, we choose models with fewer parameters, i.e. \textit{wavlm-base} and \textit{bert-base-uncased}, and run the frozen models to extract the embeddings on-the-fly during training. The resulting dimensions of audio and word embeddings are both $D_{\bm{W}}=768$.

% {\bf WavLM and BERT~} For audio and word embedding extraction, we use \textit{wavlm-base} and \textit{bert-base-uncased} to extract compact embeddings. WavLM shrinks the length of the input waveform by a factor of 320, and both WavLM and BERT outputs 768 dimensional embeddings.\\
\noindent{\bf Summarizer Transformer~} Both the subword-level summarizer and sentence-level aggregator have 768 input dimensions, which match WavLM's embedding, and are based on the same 4-layer Transformer encoder structure. As positional encodings are used in both WavLM and BERT, the summarizer Transformer opts not to use them.
For the summarizer Transformer pretraining, Adam is again used with the learning rate, $\beta_1$, and $\beta_2$ set to be $10^{-4}$, $0.9$, and $0.98$, respectively. No learning rate scheduler is involved. Early stopping engages when the validation loss is not improved for 30 epochs. $\mathcal{L}_\text{TTR}$ in eq. \eqref{eq:ttr} is used for the pretraining, while finetuning is to reduce $\mathcal{L}_\text{total}$ by jointly updating both the separator $\mathcal{S}$ and the summarizer $\mathcal{P}$. 

% \noindent{\bf Optimization~} \minje{Provide information about pretraining CTN and summarizer here.} For finetuning, we choose the Adam \cite{KingmaD2015adam} optimizer with a learning rate of $10^{-3}$. The learning rate scheduler halves the learning rate when the validation loss does not change for \minje{XXX batches, epocs or whatever}. The batch size are set to 24. \minje{Early stopping criterion.}%A learning rate of $10^{-4}$ is used for $\mathcal{P}$ pretraining.
% 
\begin{figure}[t]
  \centering
  \centerline{\includegraphics[width=\linewidth]{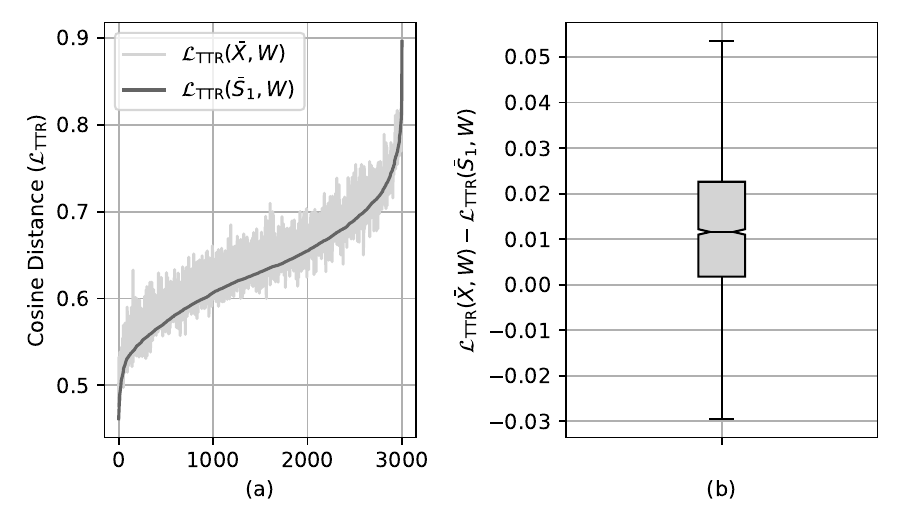}}
  \caption{Evaluation of the summarizer Transformer.}
  \label{fig:summerizer}
  % \vspace{-1pt}
\end{figure}

\section{Experimental Results and Discussion}

\subsection{Evaluation of the Summarizer Transformer}
\label{ssec:eval_summarizer}
To verify the extent to which the textual data contributes, we evaluate the summarizer Transformer using the validation set. In Figure \ref{fig:summerizer} (a), the horizontal axis represents the 3,000 utterances in the validation set, and the vertical axis stands for the cosine distance between the projected audio embeddings and subword embeddings. In this comparison, we show the embeddings from the matching source $\bar{\bm{S}}_1$ are more similar to the word embeddings $\bm W$ than the embeddings extracted from the mixtures $\bar{\bm{X}}$ are, where $\bar{\bm{X}}\leftarrow\mathcal{P}(\mathrm{SLA}(\mathcal{A}(\bm x)))$.
All the  $\mathcal{L}_{\rm TTR}(\bar{\bm{S}}_1, \bm{W})$ values are sorted ascendingly, and the sorted indices are utilized to sort the values of $\mathcal{L}_{\rm TTR}(\bar{\bm{X}}, \bm{W})$. From Figure \ref{fig:summerizer} (a), we see that the overall difference of $\mathcal{L}_{\rm TTR}(\bar{\bm{X}}, \bm{W})$ is higher than $\mathcal{L}_{\rm TTR}(\bar{\bm{S}}_1, \bm{W})$. Furthermore, Figure \ref{fig:summerizer} (b) illustrates the distribution of the difference between $\mathcal{L}_{\rm TTR}(\bar{\bm{S}}_1, \bm{W})$ and $\mathcal{L}_{\rm TTR}(\bar{\bm{X}}, \bm{W})$. The notch indicates the 95\% confidence interval, with 78.87\% of $\mathcal{L}_{\rm TTR}(\bar{\bm{X}}, \bm{W}) \geq \mathcal{L}_{\rm TTR}(\bar{\bm{S}}_1, \bm{W})$, and the mean of the difference is $1.23\times 10^{-2}$. These results indicate that the summarizer can distinguish between clean sources and mixtures based on the textual context.
% Among all samples, 78.87\% of the $\mathcal{L}_{\rm TTR}(\bar{\bar{\bm{S}}}_1, \bm{W})$ are lower than $\mathcal{L}_{\rm TTR}(\bar{\bar{\bm{S}}}_1 + \bar{\bar{\bm{S}}}_2, \bm{W})$, and the difference of their mean values is 0.0123, and their variances are 0.036. From the result, we are confident that textual data can provide some useful information in distinguishing the clean source and the mixture.

\subsection{Source Separation Performance}
In this main experiment, we compare the performance of the Conv-TasNet and SepFormer baselines, and their TTR variants. Table \ref{tab:perf} presents the evaluation improvement scores of the two separation models for each metric and task. Higher scores indicate better results. For a fair comparison, we use the pretrained checkpoints provided by the authors of both baseline models. Note that we use the SepFormer variation, which does not use dynamic mixing or pretraining.

When we apply the TTR loss to finetune the baseline models, the proposed regularization introduces performance improvements in all task variations. In Table \ref{tab:perf}, the boldface denotes the best results per subtask. Specifically, Conv-TasNet+TTR outperforms its baseline in SI-SDR by 0.31 dB in Libri2Mix-clean and 0.19 dB in Libri2Mix-noisy tasks. As for three-source mixtures, the proposed model surpasses the baseline by 0.43 dB for Libri3Mix-clean and 0.49 dB for Libri3Mix-noisy subsets, respectively.
Investigating SepFormer's performance, TTR successfully boosts SI-SDR by 1.47 dB in Libri2Mix clean separation task and by 0.83 dB in Libri2Mix noisy task. 
% However, the SIR scores showed an inconsistent trend as compared with other scores.
A similar trend can also be found in Libri3Mix tasks where SI-SDR scores was improved by 1.15 dB in Libri3Mix-clean and 1.18 dB in Libri3Mix-noisy, respectively.

% To encapsulate, since the corresponding SDR and SIR values have all been improved, this means that the TTR-SS model is able to produce separated sources with higher separation quality and less distortion than the baseline models. 
Since all SDR and STOI scores follow a similar trend as SI-SDR, we draw the conclusion that the TTR-SS models are able to produce separated sources with higher quality and intelligibility.
Additionally, it is evident that the performance enhancement derived from the SepFormer baselines surpasses that from Conv-TasNet, suggesting that the proposed regularization provides more information to learn for larger models.

\begin{table}[t]
\centering
\caption{Source separation performance. SDR and SI-SDR improvements are reported in decibel (dB), while STOI values range between 0 and 1, where 1 is the upper bound.}
\label{tab:perf}
\resizebox{0.92\columnwidth}{!}{%
\begin{tabular}{c|c|c c c}
\toprule
Task & Model ($\lambda$) & SDRi & SI-SDRi & STOI \\
\midrule
\multirow{8}{*}{\shortstack[c]{Libri2Mix\\Clean}} 
& Conv-TasNet (N/A) & 15.11 & 14.76 & 0.9311 \\
& + TTR (1.0)       & 15.42 & 15.07 & 0.9342 \\
& + TTR (0.5)       & 15.42 & 15.08 & 0.9341 \\
& + TTR (0.1)       & \textbf{15.44} & \textbf{15.10} & \textbf{0.9344} \\
\cline{2-5}
& SepFormer (N/A)   & 18.68 & 18.35 & 0.9574 \\
& + TTR (1.0)       & 20.12 & 19.82 & 0.9682 \\
& + TTR (0.5)       & 20.14 & 19.85 & 0.9681 \\
& + TTR (0.1)       & \textbf{20.17} & \textbf{19.87} & \textbf{0.9685} \\
\hline
\multirow{8}{*}{\shortstack[c]{Libri2Mix\\Noisy}} 
& Conv-TasNet (N/A) & 12.36 & 11.80 & 0.8490 \\
& + TTR (1.0)       & \textbf{12.54} & \textbf{11.99} & \textbf{0.8540} \\
& + TTR (0.5)       & 12.29 & 11.74 & 0.8482 \\
& + TTR (0.1)       & 12.47 & 11.90 & 0.8512 \\
\cline{2-5}
& SepFormer (N/A)   & 15.11 & 14.54 & 0.8949 \\
& + TTR (1.0)       & 15.99 & 15.37 & \textbf{0.9103} \\
& + TTR (0.5)       & \textbf{16.00} & \textbf{15.39} & 0.9100 \\
& + TTR (0.1)       & 15.98 & 15.36 & 0.9096 \\
\midrule
\multirow{8}{*}{\shortstack[c]{Libri3Mix\\Clean}} 
& Conv-TasNet (N/A) & 12.40 & 11.98 & 0.8365 \\
& + TTR (1.0)       & \textbf{12.82} & \textbf{12.41} & \textbf{0.8448} \\
& + TTR (0.5)       & 12.76 & 12.35 & 0.8439 \\
& + TTR (0.1)       & 12.77 & 12.35 & 0.8438 \\
\cline{2-5}
& SepFormer (N/A)   & 17.26 & 16.91 & 0.9141 \\
& + TTR (1.0)       & 18.43 & 18.06 & 0.9289 \\
& + TTR (0.5)       & \textbf{18.45} & \textbf{18.09} & \textbf{0.9292} \\
& + TTR (0.1)       & 18.39 & 18.03 & 0.9291 \\
\hline
\multirow{8}{*}{\shortstack[c]{Libri3Mix\\Noisy}} 
& Conv-TasNet (N/A) & 10.93 & 10.39 & 0.7669 \\
& + TTR (1.0)       & \textbf{11.41} & \textbf{10.88} & \textbf{0.7793} \\
& + TTR (0.5)       & 11.37 & 10.84 & 0.7776 \\
& + TTR (0.1)       & 11.35 & 10.81 & 0.7769 \\
\cline{2-5}
& SepFormer (N/A)   & 14.73 & 14.24 & 0.8489 \\
& + TTR (1.0)       & \textbf{15.94} & \textbf{15.42} & \textbf{0.8727} \\
& + TTR (0.5)       & 15.88 & 15.36 & 0.8717 \\
& + TTR (0.1)       & 15.91 & 15.39 & 0.8720 \\
\bottomrule
\end{tabular}}
\end{table}

\section{Conclusion}
We presented a novel timed text-based regularization method that leverages sentence-level semantics from a language model, enhancing speech separation performance across diverse speaker and noise environments. 
For this purpose, we introduced the SLA and summarizer Transformer to align and minimize the gap between different modalities, i.e. audio and text. 
In our experiment, we demonstrated that the SLA and summarizer Transformer effectively differentiate between mixtures and clean sources by comparing them with their respective textual representations, indicating a meaningful regularization.
Consequently, TTR enhances all evaluation metrics, particularly for SepFormer, a more complex and sizable model, at higher SNR levels.
This underscores the efficacy of the proposed TTR method, which works with the conventional PIT loss, thus improving source separation tasks. The benefit comes with no additional computational or data collection cost during the test time due to its regularization-based approach. 

\pagebreak
\section{Acknowledgement}
This material is based on work supported in part by the National Science Foundation under Grant No. 2046963. The authors appreciate Dr. Cem Sübakan for providing invaluable information regarding the development of the baseline models. Part of the work was done when the authors were with Indiana University.

\bibliographystyle{IEEEtran}
\bibliography{mjkim}

\end{document}